\documentclass[english,superscriptaddress,floatfix,twocolumn,oneside, amsmath,amssymb,amsfonts
aps,pra,floatfix, reprint]{revtex4-1}
\usepackage[english]{babel}
\usepackage[utf8x]{inputenc}
\usepackage{microtype}
\usepackage{graphicx}
\usepackage{siunitx}
\usepackage{xspace}
\usepackage[hidelinks]{hyperref}
\usepackage{bm}
\usepackage{bbold}
\usepackage{natbib}
\usepackage{csquotes}
\usepackage{physics}
\usepackage{xcolor}

\usepackage[draft]{fixme}

\usepackage[capitalise]{cleveref}

\renewcommand{\vec}[1]{\ensuremath{\bm{#1}}\xspace}

\newcommand{\cnot}{\textsc{cnot}\xspace}
\newcommand{\cz}{\textsc{cz}\xspace}
\newcommand{\notgate}{\textsc{not}\xspace}
\newcommand{\swap}{\textsc{swap}\xspace}
\newcommand{\iswap}{$i$\swap}

\newcommand{\tf}{\textsc{tensorflow}\xspace}
\newcommand{\tfq}{\textsc{tensortlow quantum}\xspace}
\newcommand{\cirq}{\textsc{cirq}\xspace}
\newcommand{\adam}{\textsc{adam}\xspace}

\newcommand{\change}[1]{\textcolor{black}{#1}}

\newcommand{\affA}{Department of Physics and Astronomy, Aarhus University, DK-8000 Aarhus C, Denmark}
\newcommand{\affB}{Kvantify Aps, DK-2300 Copenhagen S, Denmark}
\newcommand{\affC}{Department of Chemistry, Aarhus University, DK-8000 Aarhus C, Denmark}

\date{\today}
\begin{document}
	
	\title{Multiqubit state learning with entangling quantum generative adversarial networks}
	
	\author{S. E. Rasmussen}
	\email{stig@phys.au.dk}
	\affiliation{\affA}
	\affiliation{\affC}
	\affiliation{\affB}
	\author{N. T. Zinner}
	\email{zinner@phys.au.dk}
	\affiliation{\affA}
	\affiliation{\affB}
	
	\begin{abstract}
		The increasing success of classical generative adversarial networks (GANs) has inspired several quantum versions of GANs. Fully quantum mechanical applications of such quantum GANs have been limited to one- and two-qubit systems. In this paper, we investigate the entangling quantum GAN (EQ-GAN) for multiqubit learning. We show that the EQ-GAN can learn a circuit more efficiently compared with a SWAP test. We also consider the EQ-GAN for learning eigenstates that are variational quantum eigensolver (VQE)-approximated, and find that it generates excellent overlap matrix elements when learning VQE states of small molecules. However, this does not directly translate into a good estimate of the energy due to a lack of phase estimation. Finally, we consider random state learning with the EQ-GAN for up to six qubits, using different two-qubit gates, and show that it is capable of learning completely random quantum states, something which could be useful in quantum state loading.
	\end{abstract}
	
	\maketitle
	
	Generative machine learning has been highly influenced  by generative adversarial networks (GANs) \cite{Goodfellow2014}. In the past couple of years, GAN applications have increased rapidly; from generating photorealistic images, \cite{Karras2019} and improving video game resolution \cite{Wang2018} to modeling dark matter \cite{Mustafa2019} and improving astrophysical images \cite{Schawinski2017}.
	
	The idea behind GANs is that two networks play a game where one network, the generator, is trying to fool the other network, the discriminator, by producing realistic data. Thus, the generator is trained indirectly through the discriminator, hence the name adversarial. Training GANs can be formidably difficult with vanishing gradients and mode collapse \cite{Arjovsky2017} and convergence problems \cite{Salimans2016,Mescheder2018}.
	
	The success of classical GANs has inspired quantum physicists to develop a quantum version, a so-called QuGAN. This was first proposed by the authors of Refs. \cite{Dallaire-Demers2018,Lloyd2018}, who argued that a fully quantum mechanical GAN reaches its Nash equilibrium \cite{Osborne1994} when the data are reproduced correctly. This QuGAN does, however, not always converge. In certain cases, it oscillated between a finite set of states due to mode collapse, as it suffers from a non unique Nash equilibrium \cite{Niu2021}. 
	
	In the original proposal, the QuGAN was shown to be able to reproduce a controlled-\notgate (\cnot) gate \cite{Dallaire-Demers2018}. QuGANs have been used to approximate pure quantum states \cite{Benedetti2019} and the modified National Institute of Standards and Technology (MNIST) datasets \cite{Huang2021}, and a proof-of-principle experiment has shown their feasibility \cite{Hu2019}.
	QuGANs have also been developed with a quantum generator and a classical discriminator and used on classical data sets, i.e., financial modeling \cite{Zoufal2019}, discrete and continuous distributions \cite{Situ2020,Romero2019}, the Bars-and-Stripes data set \cite{Zeng2019}, and sampling particle traces \cite{Chang2021}. Furthermore, they have been discussed using a quantum generator and a SWAP test discriminator \cite{Stein2021}. 
	In this paper, we consider a fully entangling quantum GAN (EQ-GAN), i.e., generator, discriminator, and data are quantum mechanical. In its proposal, the EQ-GAN is shown to be able to generate a single-qubit gate \cite{Niu2021}.
	We apply this EQ-GAN to multiqubit states of up to six generator qubits and 12 discriminator qubits. We apply it to states found using a variational quantum eigensolver (VQE) algorithm, to see whether it can reproduce data from another quantum algorithm and to completely random quantum states. Generation of quantum states has previously been studied using classical neural networks \cite{Carleo2017,Carleo2017b,Liu2019,Hibat-Allah2020,Yamming2022}.

	The paper is structured as follows: \Cref{sec:eqgan} presents a short introduction to the EQ-GAN and the circuit architecture. In \cref{sec:results} we explain how the simulation is performed, and then in  \cref{sec:Rnd2QubitCircuit} we discuss results for random two-qubit circuits, in \cref{sec:vqeResults} we use an EQ-GAN for learning VQE states, and in \cref{sec:rndStates} we discuss results when the EQ-GAN is learning completely random multiqubit states. Finally, we present our conclusion and the outlook.
	
	\section{EQ-GAN}\label{sec:eqgan}
	
	\begin{figure*}
		\centering
		\includegraphics[scale=.7]{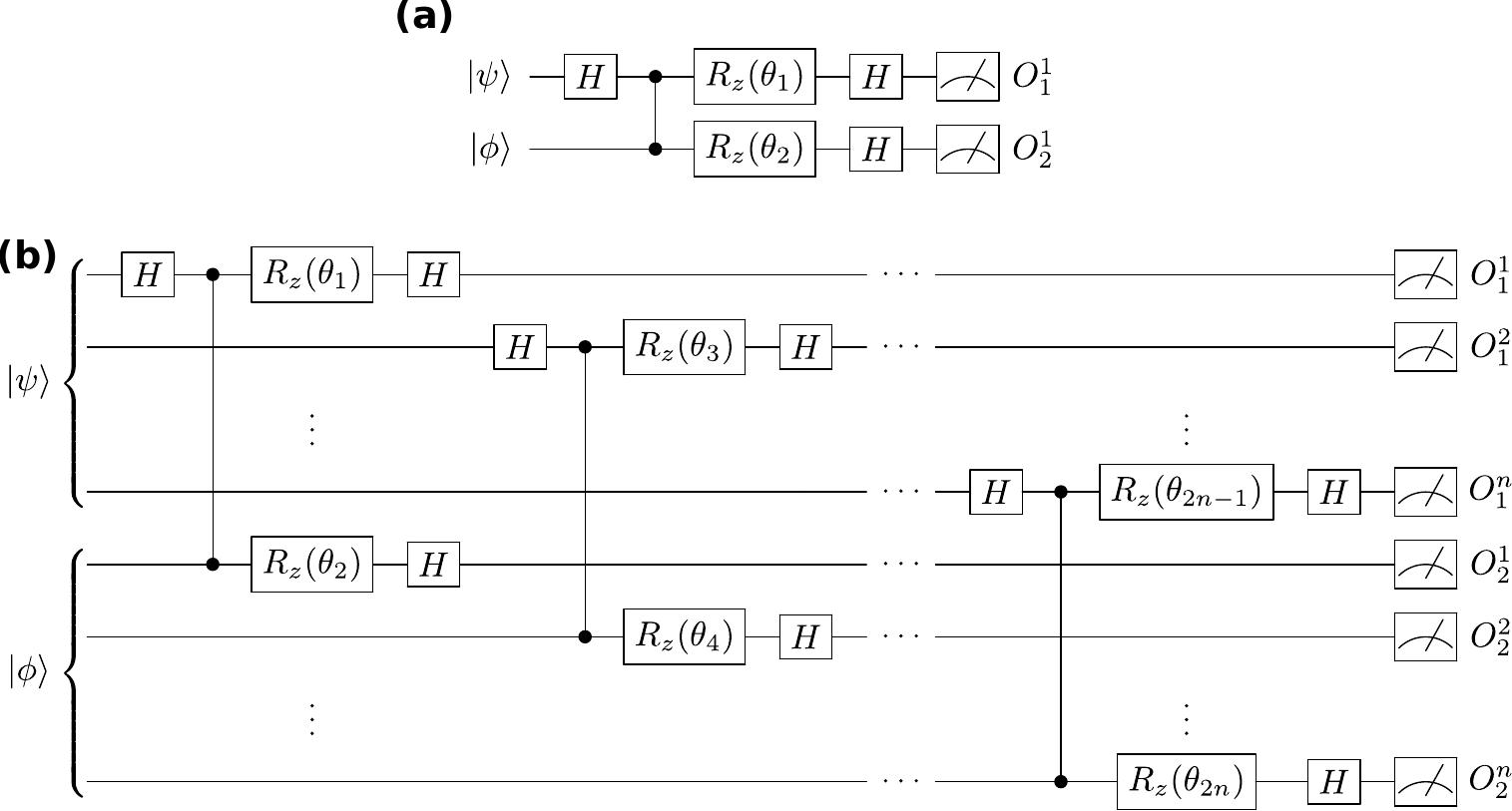}
		\caption{Parametrized (destructive) SWAP test used as discriminator in this paper. In order to obtain the original SWAP test one should replace the controlled-\textit{z} (\cz) gates by a \cnot gate, and remove the $R_z$ gates and the Hadamard gates on the qubits representing the $\ket{\psi}$. This is equivalent to setting $\theta_i = 0$ for all $i$. Measuring the state of each qubit after the SWAP test we take a bit wise \textsc{AND} operation of the output $O_1$ and $O_2$. If this yields $-1$, then the two states are identical. \textbf{(a)} The one-qubit case considered in the original proposal \cite{Niu2021}. \textbf{(b)} The multiqubit case used in this paper.}
		\label{fig:swapTest}
	\end{figure*}
	
	A classical GAN consists of two neural networks: a generator and a discriminator. The discriminator randomly receives data either from the real world, denoted as real data, or from the generator, denoted as generated or fake data. The data can be anything. The discriminator's task is to determine whether the data it receives are real or fake. On the other hand, the generator is trained to generate data to fool the discriminator. The two networks are thus adversaries: one trying to minimize the cost function and the other trying to maximize it. For a properly trained GAN, equilibrium is reached when the discriminator can only randomly guess whether the data are real or fake. \change{For the details of the cost function see \cref{app:costFunc}.}
	
	In the original proposal of a QuGAN \cite{Dallaire-Demers2018,Lloyd2018} it was proposed to exchange the neural networks with parametrized quantum circuits (PQCs) also known as quantum neural networks \cite{Gupta2002,Farhi2018,Beer2020}. Here, the discriminator becomes a quantum circuit which is then given a real quantum state or one generated from another quantum circuit.
	
	This paper considers the entangling quantum GAN (EQ-GAN) \cite{Niu2021}. In the EQ-GAN, real and fake (generated) quantum states are passed to the discriminator simultaneously such that the discriminator is allowed to entangle these. The real data could be some quantum state measured in an experiment, or it could come from some unknown quantum circuit. We denote the unitary producing this state $R$; it may or may not be dependent on a set of parameters. The fake quantum state is generated by a PQC given by some circuit \textit{Ansatz}. We denote the generator $G(\vec{\theta}_g)$ where $\vec{\theta}_d$ denotes the parameters of the generator. 
	
	The discriminator can be either untrainable as a SWAP test \cite{Burhman2001,Garcia-Escartin2013,Stein2021} or a trainable circuit capable of learning the SWAP test. The authors of Ref. \cite{Niu2021} show that the trainable discriminator outperforms the perfect SWAP test in the presence of noise. In this paper, we consider both cases and denote the discriminator $D(\vec{\theta}_d)$ where $\vec{\theta}_d$ is the discriminator's parameters. The discriminator circuit can be seen in \cref{fig:swapTest}(b) for the multiqubit case considered in this paper. It should be compared with the single-qubit case in \cref{fig:swapTest}(a) which was considered in the original proposal \cite{Niu2021}.

	\section{Results}\label{sec:results}
	
	\begin{figure}
		\centering
		\includegraphics{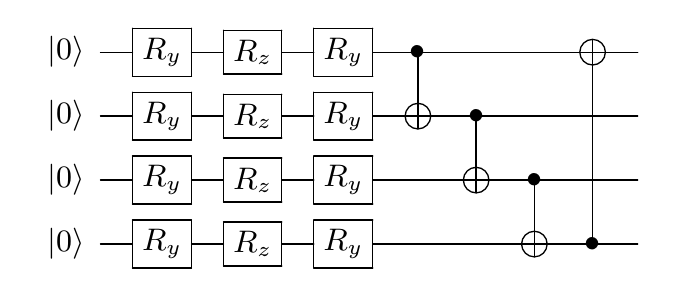}
		\caption{Example of one layer of the circuit \textit{Ansatz}, $G_l(\vec \theta_l)$, for four qubits. First, an Euler rotation is performed on each qubit, followed by a coupling using the given entangling gate. Here it is shown using fixed \cnot gates.}
		\label{fig:ParamGatesCircuit}	
	\end{figure}
	
	We simulate our EQ-GAN following the approach described in Ref. \cite{Broughton2021}. We prepare our circuits using \cirq and then add them as layers using \tfq. We compile the model using \tf and optimize it using the \adam optimizer, with learning rates of 0.01.
	We introduce noise in our system on the rotation angle of all of the gates. We do this by adding after each gate another gate with a random error.
	On the single-qubit rotations we add an error to the rotation angle sampled from a normal distribution with $\mu = 0.06$ and $\sigma = 0.02$. On the two-qubit gates we sample from a normal distribution with $\mu = 0$ and $\sigma = 0.005$. These are chosen similarly to the errors in Ref. \cite{Niu2021}.
	
	For the generator we use a layered circuit \textit{Ansatz} such that $G(\vec \theta_g) = G_L(\vec \theta_L) \cdots G_2(\vec \theta_2) G_1(\vec \theta_1)$, where each layer, $G_l(\vec \theta_l)$, consists of an Euler rotation on each qubit followed by nearest-neighbor couplings of all qubits using two-qubit gates. This means that we have $3N$ single-qubit rotations in each layer, where $N$ is the number of qubits. An example of a layer can be seen in \cref{fig:ParamGatesCircuit}, where the layer is shown with \cnot gates.
	
	We apply the EQ-GAN to several different real quantum states. First, we consider learning a simple two-qubit circuit with the same structure as the generator, then we consider states generated by a VQE algorithm, and finally, we consider completely random states.

	\subsection{Random two-qubit circuit}\label{sec:Rnd2QubitCircuit}
	
	\begin{figure}
		\centering
		\includegraphics[width=\columnwidth]{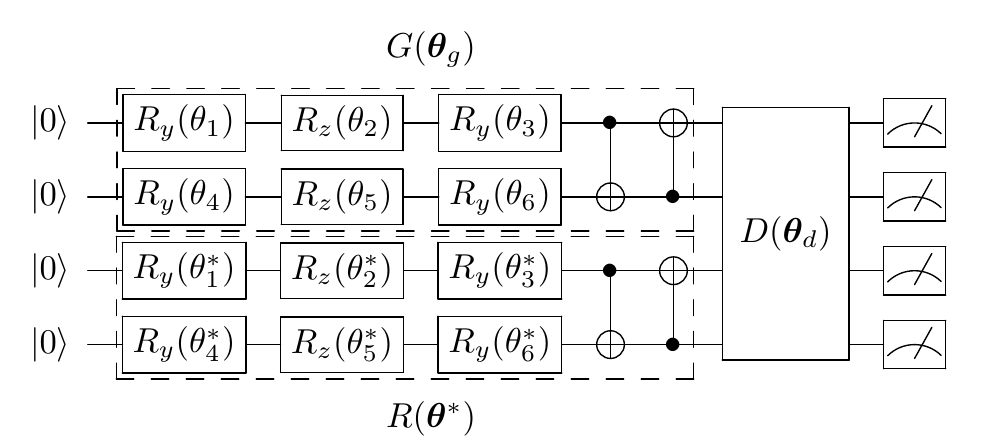}
		\caption{Circuit layout of a two-qubit EQ-GAN with a generator using the same circuit layout as the real quantum state. Note that the generator and real circuit constitute one layer of the circuit \textit{Ansatz} shown in \cref{fig:ParamGatesCircuit}.}
		\label{fig:simple2qubit}
	\end{figure}
	
	First, we consider a simple circuit with a one-layer layout identical to the generator, but with randomly chosen predefined rotation angles, $\vec{\theta}^*$. By choosing the layout of the generator to be identical to the real data, we should, in principle, be able to generate the exact same state. The full layout of the EQ-GAN can be seen in \cref{fig:simple2qubit}.
	
\begin{figure}
		\centering
		\includegraphics[width=\columnwidth]{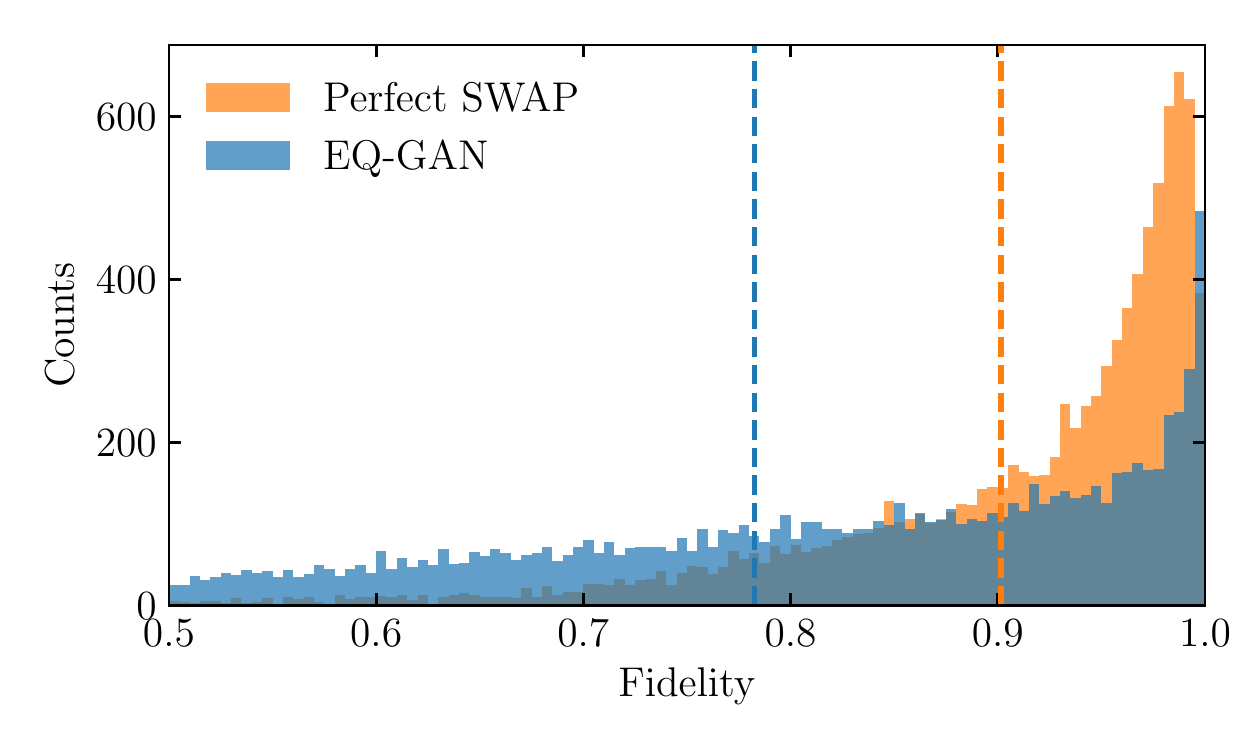}
		\caption{Distribution of $\SI{e4}{}$ EQ-GAN simulations using the simple circuit \textit{Ansatz} shown in \cref{fig:simple2qubit}. The dashed lines indicate the average fidelity of the simulations for each case.}
		\label{fig:simpleDistribution}
	\end{figure}

	We train for 80 episodes and a batch size of 4, with learning rates of 0.01 for both networks. We sample $\SI{e4}{}$ different real parameters, $\vec \theta^*$, and perform an EQ-GAN simulation on each sample for both a perfect SWAP test and an adversarial SWAP test. The distribution of fidelities, i.e., $|\bra{0}G(\vec{\theta}_g)^\dagger R(\vec{\theta}^*)\ket{0}|^2$, where $G(\vec{\theta})\ket{0}$ is the generated state and $R(\vec{\theta}^*)\ket{0}$ is the real state, can be seen in \cref{fig:simpleDistribution} where we also have plotted the average fidelity of all the simulations. We observe that the perfect swap performs better than the adversarial EQ-GAN on average. The perfect swap simulations have a larger peak and density above 0.95 and a smaller tail. However, the mode, i.e., the most frequent fidelity, of the adversarial training is larger than that of the perfect swap. The fact that the mode of the data is significantly larger than the fidelity is means that the data are more spread out as we also observe. This is a result of the difficulty in training adversarial networks. The perfect SWAP test requires no training besides the generator and thus it is easier to train on average. On the other hand, the adversarial EQ-GAN is more difficult to train as it requires both training of the generator and training of the discriminator, yielding a lower average fidelity but a large mode. This larger mode shows that the EQ-GAN has the potential to be better trained, i.e., yielding a higher maximum fidelity, compared with the perfect SWAP training. This difficulty in training GANs is well documented for the classical versions \cite{Mescheder2018}.
	
	\Cref{tab:eqgan} shows the rotation angles of the best EQ-GAN result. Despite a relatively high fidelity (above 0.9999), we observe that the rotation angles are far from identical to the real rotation angles, $\vec\theta^*$. Since it is possible to rotate around the Bloch sphere using different approaches and still end up at the same place, this is not an immediate problem. However, this is a consequence of the fact that the fidelity measure, and thus the SWAP test, only considers the amplitude of a given state and not the phases. Thus we have not generated two identical states; rather we have just generated two states with identical amplitudes. This should be kept in mind when working with EQ-GANs, as it is typical for all simulations.
	

	\begin{table}
		\centering
		\caption{Parameters used in the EQ-GAN simulation and overlap fidelity of the states returned by the circuits.  $\vec{\theta}^*$ refer to the predefined rotation angles of the real circuit in \cref{fig:simple2qubit}. $\vec\theta_p$ refer to rotation angles generated using a perfect SWAP test, and $\vec\theta_a$ refer to rotation angles generated using an adversarial SWAP test as the discriminator. All angles are in radians.}
		\label{tab:eqgan}
		\begin{tabular}{lccccccc}
			\toprule
			& Fidelity & $\theta_1$ & $\theta_2$ & $\theta_3$ & $\theta_4$ & $\theta_5$ & $\theta_6$  \\
			\hline
			$\vec{\theta}^*$ & 1 & 0.035 & 2.861 & 0.606 & 0.361 & 6.174 & 4.513 \\
			$\vec{\theta}_p$ & 0.95066 & 0.316 & 0.546 & 0.295 & 5.644 & 0.662 & 5.663 \\
			$\vec{\theta}_a$ & 0.99995 & 0.219 & 6.273 & 0.347 & 5.589 & 0.056 & 5.567 \\
			\hline
		\end{tabular}
	\end{table} 
	
	\subsection{VQE learning}\label{sec:vqeResults}
	
	We now increase the complexity of our EQ-GAN model, both in the number of qubits and in the number of layers of the generator, and by using real data.  As in the previous section, we keep the circuit architecture identical for both the real data and the generator, meaning that in principle we should be able to generate a circuit with unity fidelity. However, instead of learning random states as in Ref. \cite{Benedetti2019}, we want to learn states produced by a variational quantum eigensolver \cite{Peruzzo2014,McClean2016,Omalley2016,Kandala2017,Cao2019,Barkoutos2018,McCaskey2019,Gard2020}. In particular, we used results from an subspace-search variational quantum eigensolver (SSVQE) \cite{Nakanishi2019} simulation used for approximating small benchmark molecules The reason we use this particular variational quantum algorithm is that it can encode several states in a single circuit, thus encoding more information in the circuit. The SSVQE model has the same layered circuit \textit{Ansatz} as the EQ-GAN, i.e., the one in \cref{fig:ParamGatesCircuit}, with the same number of layers. Simulating the SSVQE algorithm for the two lowest eigenstates of a molecular Hamiltonian produces a set of parameters $\vec\theta_\text{VQE}$. The real data are thus generated by the circuit represented by $R(\vec\theta_\text{VQE})$.
	In other words, we are trying to learn approximate eigenstates of a given Hamiltonian, without knowing the actual Hamiltonian. 
	
	When the circuit we are trying to learn becomes bigger we need more data to make the training more efficient. Therefore we generate more training data from each VQE-eigenstate by sampling 100 sets of parameters from a normal distribution with $\vec \theta_\text{VQE}$ as the mean value of a standard deviation of 0.01. Other than that we keep the hyperparameters identical to the ones used in \cref{sec:Rnd2QubitCircuit}. We use the circuit \textit{Ansatz} shown in \cref{fig:ParamGatesCircuit}.
	
\begin{figure}
		\centering
		\includegraphics[width=\columnwidth]{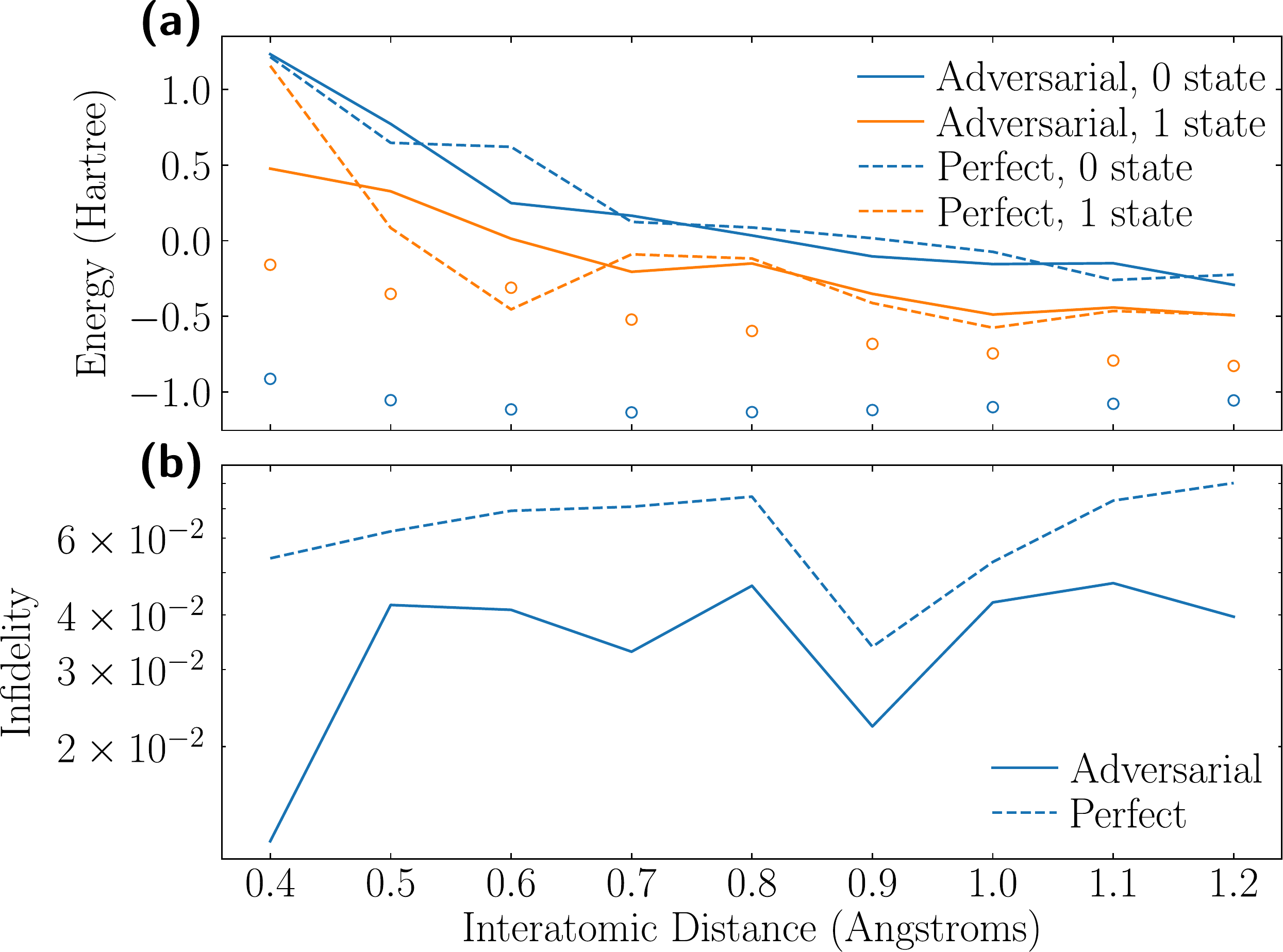}
		\caption{Results of the EQ-GAN learning on VQE data for H$_2$ with three layers of the circuit \textit{Ansatz} shown in \cref{fig:ParamGatesCircuit}. \textbf{\textup{(a)}} Predicted energy of the EQ-GAN algorithm (lines) and SSVQE (circles). Solid lines indicate that adversarial training was used, while the dashed line indicate that the perfect SWAP test was used. \textbf{\textup{(b)}} Infidelity of the state generated by the EQ-GAN, $\ket{\textup{GAN}} = G(\vec{\theta})\ket{0}$, with the state generated by the VQE, $\ket{\textup{VQE}} = R (\vec{\theta}_\textup{VQE})\ket{0}$.}
		\label{fig:H2_GAN}
	\end{figure}

	In \cref{fig:H2_GAN} we present results for simulations of H$_2$. In \cref{app:additionalRes}, we show results for LiH and BeH$_2$. The active space of H$_2$ and LiH is encoded using four qubits, while the active space of BeH$_2$ is encoded using six qubits. We simulate the molecules for different atomic bond lengths performing a simulation for each $\SI{0.1}{\angstrom}$. We plot the energy of the VQE eigenstates and the energy of the generated states. We also plot the infidelity, $1 - |\bra{0}G(\vec{\theta}_g)^\dagger R(\vec{\theta}_\text{VQE})\ket{0}|^2$, as a function of the bond length. Starting from the infidelity in \cref{fig:H2_GAN}(b), we see that the adversarial training outperforms the perfect SWAP test, and in general, we find an infidelity on the order of $10^{-2}$ equivalent to a fidelity around 0.99. Despite this quite good infidelity, we find that the energy of the generated states in \cref{fig:H2_GAN}(a) is nowhere near the SSVQE energies. They are consistently above the SSVQE energies for both the ground state and the first excited state, no matter which discriminator we use. This is due to the lack of phase estimation of the EQ-GAN. As pointed out, the lack of phase estimation is a limitation of the EQ-GAN. However, it is not likely to be something that can be mended as that may imply a violation of the no-cloning theorem.
	The phases of an eigenstate are important, and as the EQ-GAN does not approximate these the resulting energies are poorly reproduced, despite the overlap with the VQE-eigenstates. For LiH and BeH$_2$ we find similar results, see \cref{app:additionalRes}.

	\subsection{Random state learning}\label{sec:rndStates}

	Finally, we consider how efficiently an EQ-GAN can learn random states. Thus $R$ is now a completely random unitary, which is not known to the generator circuit. This means that there is no guarantee that the generator will be able to reproduce $R$ efficiently. Since we use the circuit \textit{Ansatz} as in \cref{fig:ParamGatesCircuit}, which only employs nearest-neighbor interactions, it is expected that the fidelity will decrease as the number of qubits increases. Instead of just using \cnot gates as the entangling operation we also simulate the EQ-GAN using imaginary \swap (\iswap) and \cz gates and their parametrized equivalents \cite{Rasmussen2022}. Though it may seem like an advantage that the parametrized two-qubit gates have more parameters, it is not. This is because we have already saturated the number of parameters as discussed in Ref. \cite{Rasmussen2020}.
	
	We keep the training parameters as in the previous sections and sample 200 random states. For each random state, we sample 100 training states using the same approach as in \cref{sec:vqeResults}, i.e., from a normal distribution with the coefficients of the original state as the mean value and a standard deviation of 0.01.

	\begin{figure}[ht]
		\centering
		\includegraphics[width=\columnwidth]{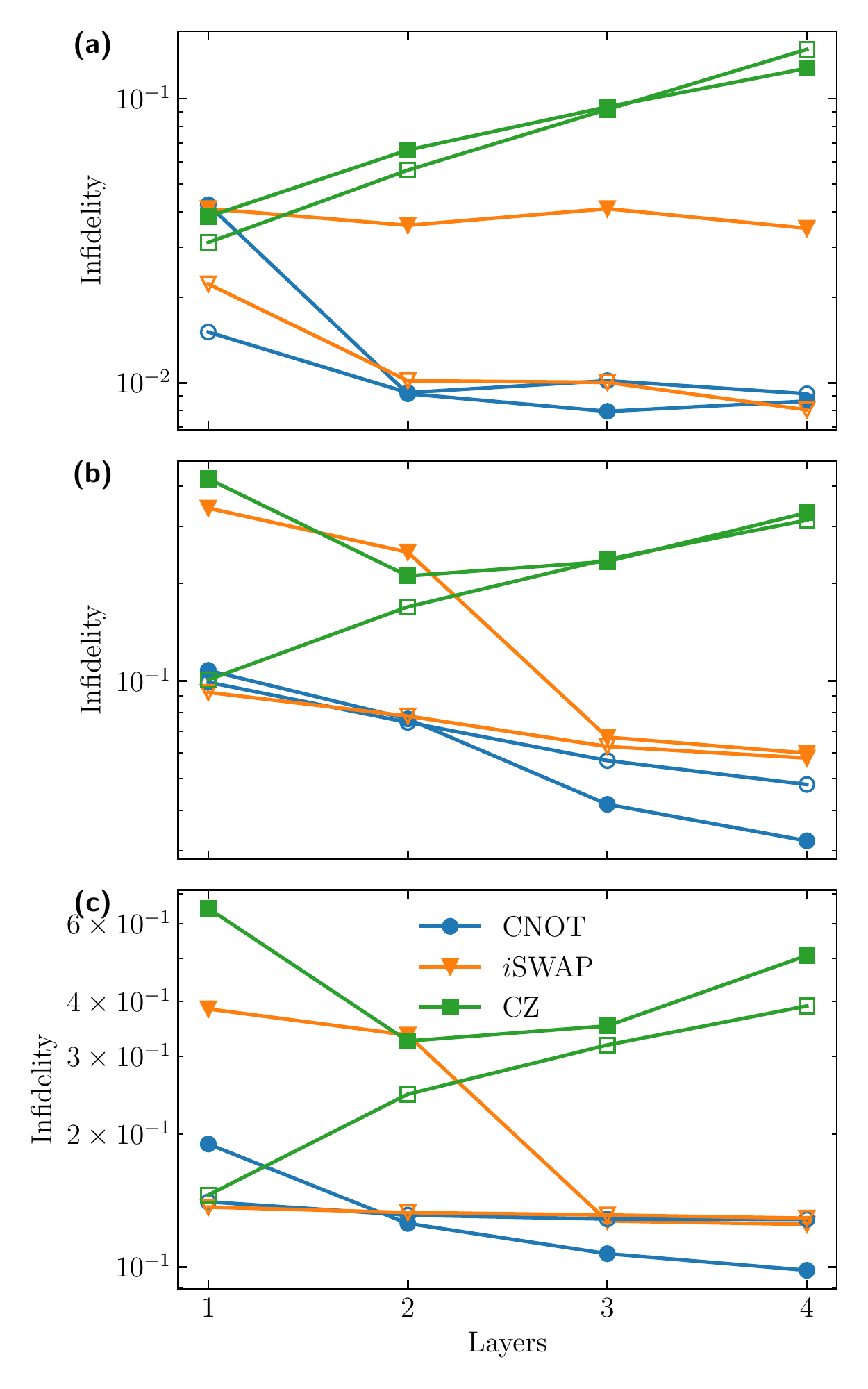}
		\caption{Average infidelity of the generated states for EQ-GAN when trying to learn random states. The sizes of the generated states are as follows: \textbf{(a)} two qubits, \textbf{(b)} four qubits, and \textbf{(c)} six qubits. Solid symbols indicate the fixed version of the two-qubit gate, while open symbols indicate parametrized versions of the two-qubit gate.}
		\label{fig:RandomStateFidel}
	\end{figure}

	We present the average infidelity of the generated states in \cref{fig:RandomStateFidel}. For the two-qubit case in \cref{fig:RandomStateFidel}(a), we observe an average infidelity of 0.1 for the worst cases and below 0.01 for the best cases. The \cz gate and its parameterize version do not perform very well, and they actually perform worse for increasing layers. This seems to be the case also for four and six qubits. Thus we conclude that the \cz gate is the least suited two-qubit gate for generative learning. This is likely because the \cz gate is diagonal in the computational basis.
	
	The \iswap gate performs better than the \cz gate, however, in the two-qubit case, its performance is approximately constant at 0.04 as the number of layers increases. The parametrized version performs better close to 0.01 on average. The \cnot gate also yields an infidelity around 0.01 on average for more than two layers, indicating that the maximum average fidelity is likely reached for this circuit \textit{Ansatz}. For one layer we see a slight advantage of the parametrized version of the \cnot gate compared with the usually fixed gate.
	
	In the four- and six-qubit cases [\cref{fig:RandomStateFidel}(b) and (c)], we see an increased minimum infidelity, since we now only have nearest-neighbor interactions. For the four-qubit case, we find the lowest infidelity at 0.02, whereas the six-qubit case does not perform better than 0.1. This increase in infidelity could be mended by changing the circuit architecture; see e.g., Ref. \cite{Sim2019} for a discussion of different circuit architectures.
	Again the \cnot gate performs best, but the infidelity is decreasing as the number of layers is increasing. The fixed and parametrized \cnot gates perform approximately identically. For the \iswap gate the parametrized version outperforms the fixed version for one or two layers while they perform identically for more layers. This indicates the advantages of using parametrized two-qubit gates instead of just increasing the number of layers. This could be advantageous in near-term devices \cite{Rasmussen2022}.

	\section{Conclusion}\label{sec:conclusion}
	
	In this paper, we have considered the EQ-GAN and its ability to learn different unitaries. 	
	We have shown that the EQ-GAN can perform better than the SWAP test, but it is more difficult to train. The perfect SWAP test yields a higher average fidelity, however, the EQ-GAN yields a higher maximum fidelity. 
	We also discuss that despite this high fidelity the trained parameters are not necessarily the same as the real parameters used to define the state. This is because the EQ-GAN lacks phase estimation of the generated states. 
	We show that this may be a problem, e.g., when trying to learn eigenstates of a Hamiltonian. In such a case, even with high fidelity, the eigenvalues of the generated states can be far from the target eigenvalues. Nonetheless, EQ-GANs can still be useful if phase estimation is not required, for instance, if we are looking for the probability of finding the amplitude of different basis states in a given quantum states. 
	
	Finally, we discussed random state learning with a layered nearest-neighbor circuit \textit{Ansatz}. We find that the \cnot gate performs best and that the parametrized versions of two-qubit gates outperform the fixed versions for a few layers. We also find that the fidelity decreases as the number of qubits in the state we want to generate increases. This could be mended by changing the circuit \textit{Ansatz} to increase the qubit interactions beyond nearest neighbor. The results could also be improved by tuning the hyperparameters of the model. In general, hyperparameter tuning is a large issue in classical generative models \cite{Salimans2016}, and it is therefore expected to be of similar difficulty for quantum generative models.
	
	Possible application of EQ-GANs could be to test how well various circuit \textit{Ans\"{a}tze} can approximate sets of random states. Our results also show that EQ-GANs could be useful for quantum state loading or distribution loading \cite{Zoufal2019,Stamatopoulos2020}.

	\begin{acknowledgments}
		The authors thank T. Bækkegaard for discussions on different aspects of the work.
		This work is supported by the Danish Council for Independent Research.
	\end{acknowledgments}

\appendix

\section{Cost function of EQ-GAN}\label{app:costFunc}

A classical GAN consists of two neural networks, a generative network, $G(\vec\theta_g, \vec z)$ and a discriminative network, $D(\vec \theta_d, \vec z)$. The generator maps a vector, $\vec z$, sampled randomly from a given distribution to a data example, $G(\vec\theta_g, \vec z)$. We call this the fake data. The discriminator takes a sample $\vec x$ and returns the probability, $D(\vec\theta_d, \vec x)$, for the data being real or fake, i.e., generated by the generator network. 

This creates a minimax optimization problem, where we are alternating between improving the discriminator's ability to distinguish between real and fake data and improving the generator's ability to trick the discriminator. In general, a classical GAN is trained by solving $\min_{\vec\theta_g}\max_{\vec \theta_d} V(\vec\theta_d \vec\theta_g)$, where the cost function is given as
\begin{equation}\label{eq:classicalGAN}
\begin{aligned}
V(\vec\theta_d, \vec\theta_g) =& \mathbb{E}_{x\sim p_\text{data}(x)}[\log D(\vec\theta_d, x)]\\
+& \mathbb{E}_{z\sim p_0(z)}\{\log[1 - D(\vec\theta_d, G(\vec\theta_g, z))]\},
\end{aligned}
\end{equation}
where $\mathbb{E}_{x\sim p_\text{data}(x)}$ and $\mathbb{E}_{z\sim p_0(z)}$ represent the expectation over the distributions $p_\text{data}$ and $p_0$ respectively. If the two neural networks approach the space of arbitrary functions, then the global optimum exists and is uniquely determined as $p_g(z) = p_\text{data}(z)$.

The first proposal of a QuGAN \cite{Lloyd2018,Dallaire-Demers2018} defined the generative network as a quantum circuit, $G(\vec\theta_g)$, that outputs the state (here represented as a density matrix) $\rho = G(\vec\theta_g) \rho_0 G^\dagger(\vec\theta_g)$ from the initial state $\rho_0$. The discriminator then takes a real state, $\sigma$, or a fake state, $\rho$, as its input $\rho_\text{in}$, and performs a positive operator-valued measurement (POVM) defined by $T$. The outcome of the POVM determines the probability of the state being real, $D(\vec\theta_d, \rho_\text{in}) = \tr [T\rho_\text{in}]$. This means that the QuGAN is trained by solving the minimax problem 
\begin{equation}
	\min_{\vec\theta_g}\max_{T} V(\vec\theta_g, T)  = \min_{\vec\theta_g}\max_{T}(\tr [T\sigma] - \tr[T\rho(\vec\theta_g)]).
\end{equation}
However, this might not converge to the desired Nash equilibrium, leading to oscillations of the generator and discriminator, as shown in Ref. \cite{Niu2021}.

To fix this problem, the authors of Ref. \cite{Niu2021} propose to let the discriminator evaluate the true data, $\sigma$, and the fake data, $\rho(\vec\theta_g)$, simultaneously, as seen in \cref{fig:simple2qubit}. They define a minimax cost function similar to the classical GAN by \cref{eq:classicalGAN}:
\begin{equation}\label{eq:EQGANcost}
	V(\vec \theta_g, \vec \theta_d) = 1 - D(\vec \theta_d, \rho(\vec\theta_g)),
\end{equation}
where $D(\vec \theta_d, \rho(\vec\theta_g))$ is the parametrized SWAP test as seen in \cref{fig:swapTest} and $\rho(\vec\theta_g)$ is the generated state. \Cref{eq:EQGANcost} defines the cost function used for the EQ-GAN considered in this paper.

\section{Additional results}\label{app:additionalRes}

\begin{figure}[ht]
	\centering
	\includegraphics[width=\columnwidth]{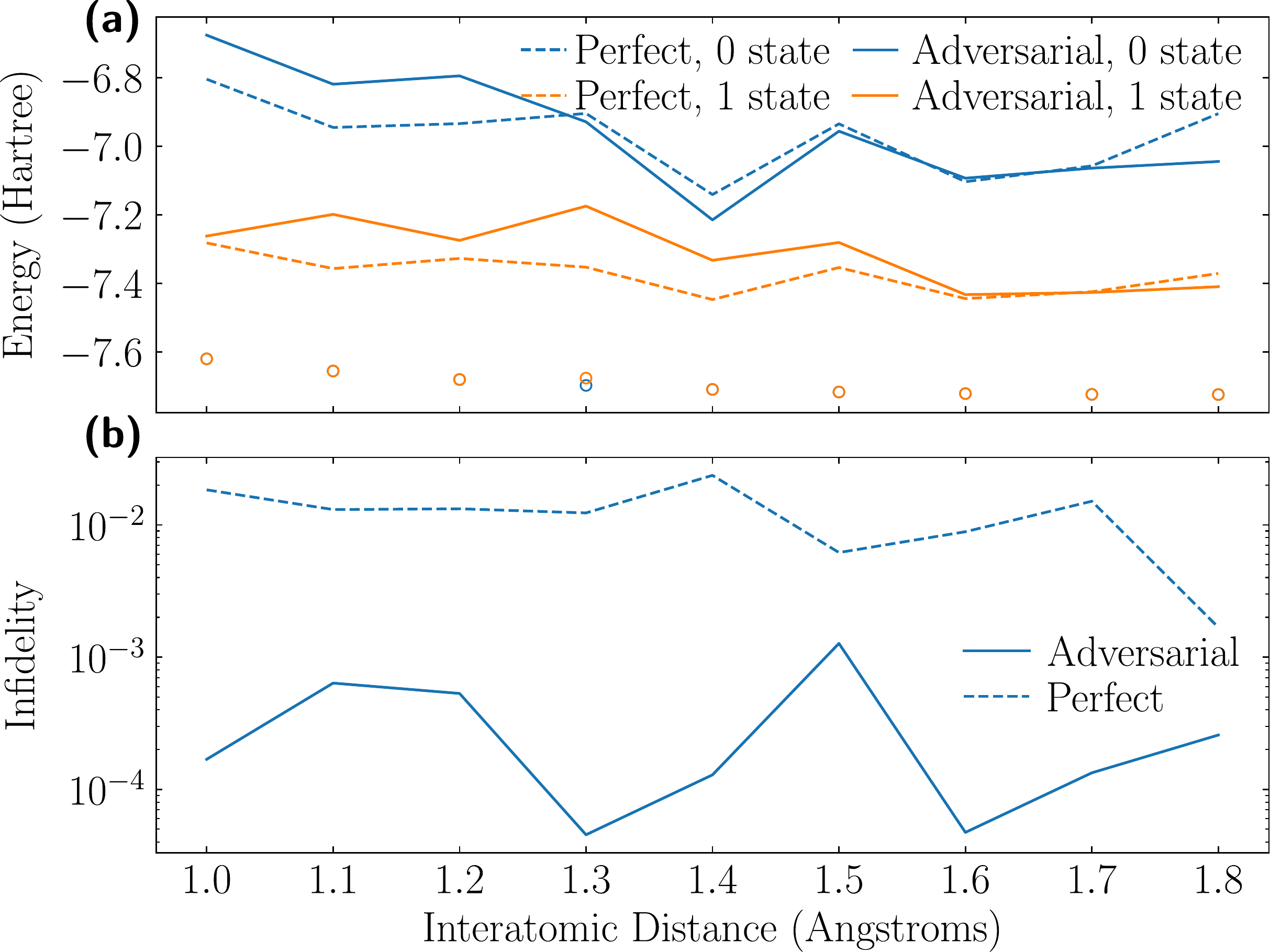}
	\caption{Results of the EQ-GAN learning on VQE data for LiH with one layer. \textbf{\textup{(a)}} Predicted energy of the EQ-GAN algorithm (lines) and SSVQE (circles). Solid lines indicate that the adversarial training was used, while the dashed line indicate that the perfect SWAP test was used. \textbf{\textup{(b)}} Infidelity of the state generated by the EQ-GAN, $\ket{\textup{GAN}} = G(\vec{\theta})\ket{0}$, with the state generated by the VQE, $\ket{\textup{VQE}} = R (\vec{\theta}_\textup{VQE})\ket{0}$.}
	\label{fig:LiH}
\end{figure}

\begin{figure}[ht]
	\centering
	\includegraphics[width=\columnwidth]{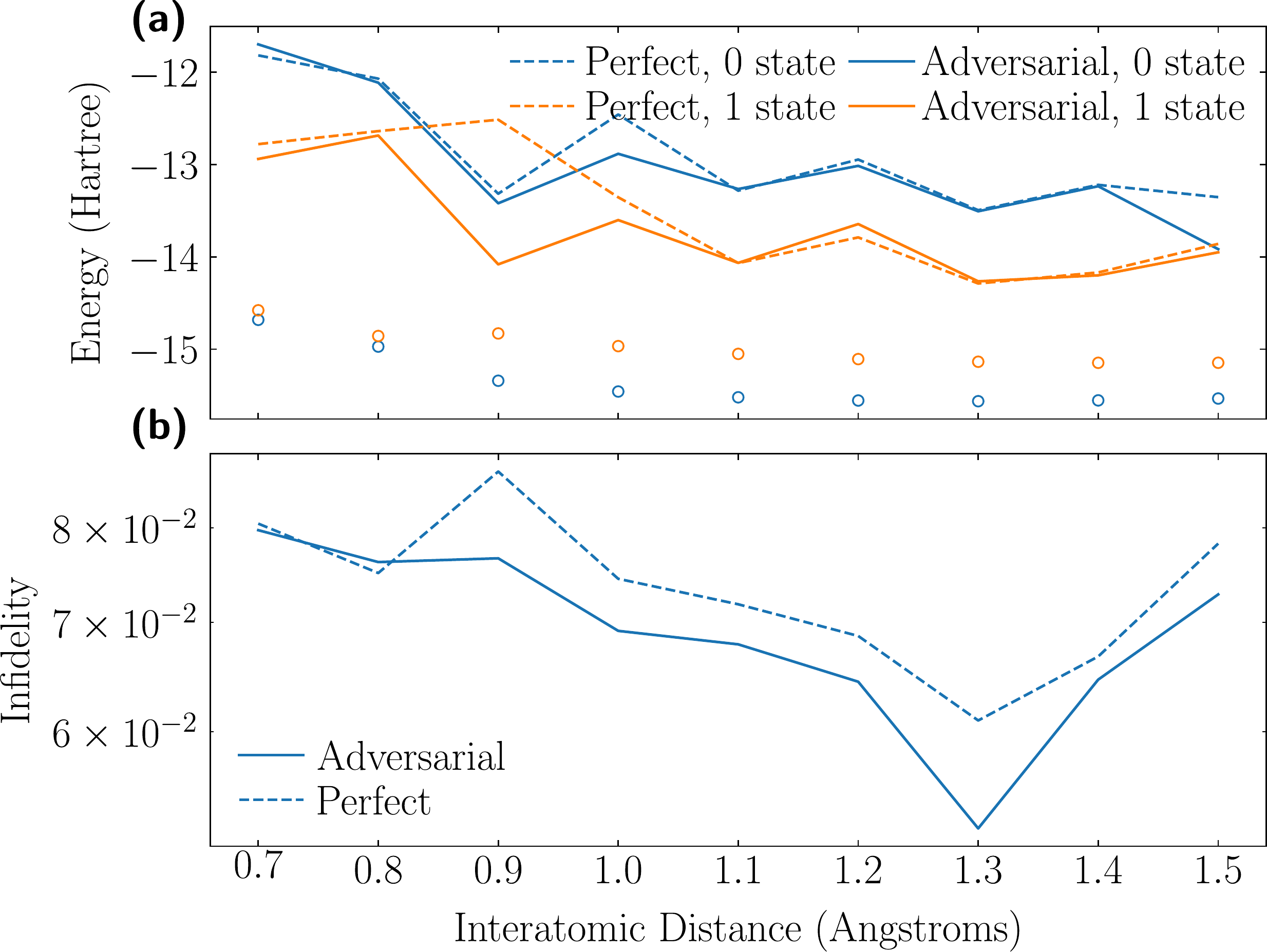}
	\caption{Results of the EQ-GAN learning on VQE data for BeH$_2$ with three layers. \textbf{\textup{(a)}} Predicted energy of the EQ-GAN algorithm (lines) and SSVQE (circles). Solid lines indicate that the adversarial training was used, while the dashed lines indicate that the perfect SWAP test was used. \textbf{\textup{(b)}} Infidelity of the state generated by the EQ-GAN, $\ket{\textup{GAN}} = G(\vec{\theta})\ket{0}$, with the state generated by the VQE, $\ket{\textup{VQE}} = R (\vec{\theta}_\textup{VQE})\ket{0}$.}
	\label{fig:BeH2}
\end{figure}

Here we present some additional results. We present simulations for LiH with one layer (\cref{fig:LiH}) and BeH$_2$ with three layers (\cref{fig:BeH2}).

\clearpage
\bibliography{QuGANBib}

\end{document}